\documentclass[conference,letter]{IEEEtran}
\usepackage{cite}
\usepackage{epsfig}
\usepackage{amssymb}
\usepackage{array}
\newcolumntype{C}[1]{>{\centering\let\newline\\\arraybackslash\hspace{0pt}}m{#1}}
\usepackage[cmex10]{amsmath, nccmath}
\usepackage{mathtools}
\usepackage{upgreek}
\usepackage{graphicx}
\usepackage{enumitem}
\usepackage{bm}
\newtheorem{theorem}{Theorem}

\newtheorem{definition}{Definition}

\newcommand{\E}[1]{\mathbb E\left[#1\right]}
\newcommand{\Prob}[1]{\mathbb P\left[#1\right]}

\newcommand{\mc}[1]{\mathcal{#1}}

\newcommand{\thmref}[1]{Theorem~\ref{#1}}

\newcommand{\figref}[1]{Fig.~\ref{#1}}

\DeclareMathOperator*{\argmax}{arg\,max}

\usepackage{url}
\usepackage{cases}
\usepackage{ifthen}
\newif\ifshowtodo
\showtodotrue
\renewcommand{\IEEEproofindentspace}{1\parindent}

\newcommand{\VersionLength}{long}
\providecommand{\ver}{\ifthenelse{\equal{\VersionLength}{long}}}
\allowdisplaybreaks
\begin{document}
\title{Variable-Length Codes with Bursty Feedback }
\author{James Y. Chen, Recep Can Yavas, Victoria Kostina
\thanks{J. Y.~Chen, and V.~Kostina are with the Department of Electrical Engineering, California Institute of Technology, Pasadena, CA~91125, USA (e-mail: jychen, vkostina@caltech.edu). R. C.~Yavas (e-mail: ryavas@caltech.edu), was formerly with the Department of Electrical Engineering, California Institute of Technology, and is now with CNRS@CREATE, 1 Create Way, \#08-01 Create Tower, Singapore 138602. This work is supported in part by the National Science Foundation (NSF) under grants CCF-1751356 and CCF-1956386 and the National Research Foundation, Prime Minister’s Office, Singapore under its CREATE programme.}}
\IEEEoverridecommandlockouts
\maketitle 
\begin{abstract}
We study variable-length codes for point-to-point discrete memoryless channels with noiseless unlimited-rate feedback that occurs in $L$ bursts. We term such codes variable-length bursty-feedback (VLBF) codes. Unlike classical codes with feedback after each transmitted code symbol, bursty feedback fits better with protocols that employ sparse feedback after a packet is sent and also with half-duplex end devices that cannot transmit and listen to the channel at the same time. We present a novel non-asymptotic achievability bound for VLBF codes with $L$ bursts of feedback over any discrete memoryless channel. We numerically evaluate the bound over the binary symmetric channel (BSC). We perform optimization over the time instances at which feedback occurs for both our own bound and Yavas \textit{et al.}'s non-asymptotic achievability bound for variable-length stop-feedback (VLSF) codes, where only a single bit is sent at each feedback instance. Our results demonstrate the advantages of richer feedback: VLBF codes significantly outperform VLSF codes at short blocklengths, especially as the error probability $\epsilon$ decreases. Remarkably, for BSC(0.11) and error probability $10^{-10}$, our VLBF code with $L=5$ and expected decoding time $N\leq 400$ outperforms the achievability bound given by Polyanskiy \emph{et al.} for VLSF codes with $L=\infty$, and our VLBF code with $L=3$.
\end{abstract}
\section{Introduction}
Many communication systems allow for the receiver to send feedback to the transmitter through a separate feedback channel. While such feedback is unable to increase the capacity of a memoryless channel \cite{shannon1956zero}, it aids in the construction of coding schemes \cite{horstein,schalkwijk,shayevitz,LiElGamal2015,naghshvar2015EJS,yang2022SED,guo2022reliability}. For instance, feedback allows for the construction of variable-length feedback (VLF) codes, in which the decoding time is a function of the feedback and the received transmission. Feedback also improves the higher-order terms of achievable rate; notably, Wagner \textit{et al.} \cite{wagner2020} show that feedback can improve the second-order achievable rate of fixed-length codes for discrete memoryless channels (DMCs) satisfying certain properties. If the rate of transmission is strictly less than the channel capacity, the associated error probability decays exponentially with blocklength following a rate-dependent exponent called the error exponent \cite{Gallager1965}. Burnashev \cite{burnashev1976data} shows that feedback improves the achievable error exponent and constructs a VLF code that achieves the optimal error exponent for feedback codes over a DMC.

While error exponent analyses \cite{Gallager1965,burnashev1976data} shed light on how fast the capacity can be approached with increasing blocklength, approximating the error probability by the error exponent might not be accurate at shorter blocklengths since such an approximation usually ignores sub-exponential terms. Polyanskiy \emph{et al.} \cite{polyanskiy2011feedback} formalize VLF codes in the non-asymptotic regime and demonstrate that at short blocklengths, VLF codes can achieve much higher rates compared to their fixed-length counterparts. They also introduce variable-length stop-feedback (VLSF) codes, which form a subset of VLF codes. In VLF codes, there is no restriction on the content of the feedback signal. In VLSF codes, the receiver sends back a single bit of noiseless feedback at each feedback instance, and feedback is used only to inform the transmitter whether a decoding decision has been made rather than to aid in the formation of future codeword symbols. Polyanskiy \emph{et al.} show that for a fixed error probability $\epsilon \in (0, 1)$, the maximum rates achievable within the classes of VLSF codes and VLF codes differ by at most $O\left(\frac{\log N}{N}\right)$, where $N$ is the expected decoding time. For the binary symmetric channel (BSC), Yang \emph{et al.} \cite[Th.~7]{yang2022SED} improve upon Polyanskiy \emph{et al.}'s VLF achievability bound by using the small-enough-difference (SED) encoder introduced in \cite{naghshvar}; their analysis refines the non-asymptotic bound in \cite{naghshvar}. Yang \emph{et al.} also extend the SED encoder to the class of binary asymmetric channels. Guo \emph{et al.} \cite{guo2022reliability} develop an instantaneous SED encoder for binary-input DMCs where the message bits stream into the encoder during transmission.

The codes introduced in \cite{burnashev1976data,polyanskiy2011feedback, yang2022SED, naghshvar} feed back after every channel use, which can be impractical due to increased power consumption and issues with half-duplex devices; so, feedback is often sparse in practice. In automatic repeat request codes\cite{arq}, the receiver feeds back a bit after receiving a transmission to report either an acknowledgment of a received message or an erasure. Forney \cite{Forney1968} derives bounds on feedback codes in which the receiver has the opportunity to report an erasure after it has received a full transmission, but cannot confidently declare a single message. Forney \cite{Forney1968} then converts these codes into VLSF codes by repeating the process until no erasure is reported and derives an achievable error exponent. VLSF codes with less frequent feedback are presented in \cite{kim2015VLF, Vakilinia2016, heidarzadeh2019Systematic, yavas2021VLF, Yang2022}. Kim \emph{et al.} \cite{kim2015VLF} consider a VLSF code with $L$ periodic feedback instances.
Vakilinia \emph{et al.} \cite{Vakilinia2016} are the first to realize that the schedule of feedback instances can be optimized; they develop a sequential procedure to do so and apply it to non-binary low-density parity-check codes. Heidarzadeh \emph{et al.} \cite{heidarzadeh2019Systematic} extend the sequential procedure in \cite{Vakilinia2016} to a scenario with a constraint on the feedback rate and apply it to erasure channels.
Yavas \textit{et al.} \cite{yavas2021VLF} derive novel non-asymptotic achievability bounds for VLSF codes with a limited number of stop-feedback instances, optimize the feedback schedule, and provide a second-order asymptotic analysis of the optimized system. 
They show that for the Gaussian channel under a maximal power constraint, the achievability bound for the VLSF code with as few as four feedback instances is close to Polyanskiy \emph{et al.}'s achievability bound \cite{polyanskiy2011feedback} with $L=\infty$. Yang \emph{et al.} \cite{Yang2022} develop numerical tools using Edgeworth and Petrov expansions and use the sequential procedure of \cite{Vakilinia2016} to optimize the decoding times given by the non-asymptotic bound in \cite{yavas2021VLF}. 

In the two-phase coding schemes studied in \cite{YamamotoItoh,LalithaAlmostFixedLength,NakiboğluZheng2012}, each feedback instance can contain multiple symbols. Yamamoto and Itoh \cite{YamamotoItoh} present a two-phase coding scheme that alternates between communication and confirmation phases. During a communication phase, the transmitter sends its intended transmission through the channel; the receiver then produces an estimate of the intended message and feeds this estimate back to the transmitter. A confirmation phase then starts, during which the transmitter sends a control sequence either confirming or rejecting the receiver's estimate; the receiver then estimates the noisy version of the control sequence via a binary hypothesis test and feeds its result back to the transmitter. If the estimate is rejected, the process restarts and is repeated anew until the transmitter eventually confirms the receiver's estimate. The communication and confirmation phases are periodic with durations of $\alpha n$ and $(1-\alpha) n$ channel uses, where $\alpha = \frac{R}{C} \in (0, 1)$, $R$ is the rate, $C$ is the channel capacity, and $n$ is the combined length. Yamamoto and Itoh's scheme achieves Burnashev's optimal error exponent \cite{burnashev1976data} despite not having feedback after each channel transmission.
A variant of Yamamoto and Itoh's coding scheme with three instances of feedback is presented in \cite{LalithaAlmostFixedLength}; if the end of the second longer communication phase is reached, the receiver accepts its current estimate there without confirmation from the transmitter. Transmission reaches the second communication phase with a vanishing probability, which allows the code to achieve Burnashev's optimal error exponent.

In this work, we introduce variable-length bursty feedback (VLBF) codes, which form a subset of VLF codes with a limited number of instances $L$ of unlimited-rate feedback. We derive the first non-asymptotic achievability bound for VLBF codes and present numerical results based on this bound. We demonstrate that at short blocklength, VLBF codes can outperform the sparse VLSF scheme in \cite{yavas2021VLF} with the same number of feedback instances. We also show that at low error tolerances, VLBF codes with $L=3$ can outperform the achievability bound in \cite{polyanskiy2011feedback} for VLSF codes with $L=\infty$. Our VLBF code alternates between communication and confirmation phases with a burst of feedback after each phase, similarly to Yamamoto and Itoh's scheme \cite{YamamotoItoh}. However, unlike \cite{YamamotoItoh}, we do not discard the symbols from the previous communication phases; while such forgetfulness causes no performance loss in the achievable error exponent \cite{YamamotoItoh,LalithaAlmostFixedLength}, it would significantly increase the expected decoding time in our non-asymptotic achievability bound. We also do not assume that the feedback instances are periodic as in \cite{YamamotoItoh}, but rather optimize their schedule as in \cite{Vakilinia2016, yavas2021VLF, Yang2022}.

The feedback codes discussed above are classified in Table~\ref{feedback table}. The codes in \cite{Vakilinia2016, LalithaAlmostFixedLength,NakiboğluZheng2012} fall into the same category as our VLBF codes, as they have $1<L<\infty$ feedback instances and allow for multiple symbols in one feedback instance. Vakilinia \emph{et al.} \cite{Vakilinia2016} optimize feedback instances for a binary-input additive white Gaussian noise channel using a Gaussian approximation; our bound applies to general DMCs and does not use approximations. The codes in \cite{LalithaAlmostFixedLength,NakiboğluZheng2012} alternate between communication and confirmation phases with a finite number of feedback bursts, just like our VLBF code; however, we focus on the non-asymptotic regime rather than the error exponent regime, and we optimize the schedule of feedback bursts.

\begin{table}[!htbp]
\centering
\label{feedback table}
\vspace{-0.3cm}
\caption{Classification of feedback codes}
\vspace{-0.2cm}
\begin{tabular}{ |C{2.2cm}| C{0.8cm}| C{2.5cm}|C{1.7cm}|} 
\hline
& \multicolumn{3}{c|}{Max number of feedback instances $L$  } \\
\hline
Max bits per feedback instance &  $L=1$ & $1<L<\infty$ & $L=\infty$\\ 
\hline
1 & -- & \cite{yavas2021VLF, Forney1968, kim2015VLF, heidarzadeh2019Systematic, Vakilinia2016, Yang2022} & \cite{arq,horstein,shayevitz,LiElGamal2015,naghshvar2015EJS,yang2022SED,guo2022reliability,polyanskiy2011feedback}\\
\hline
$>1$ & \cite{wagner2020} & VLBF codes, \cite{schalkwijk, NakiboğluZheng2012, LalithaAlmostFixedLength} & \cite{burnashev1976data, yang2022SED, naghshvar, YamamotoItoh} \\ 
\hline
\end{tabular}
\vspace{-0.2cm}
\end{table}

In the rest of this paper, Section~\ref{sec:problemstatement} formalizes VLBF codes. Section~\ref{sec:main} introduces a novel non-asymptotic achievability bound for VLBF codes. Section~\ref{sec:numer} presents numerical results.
\section{Problem Statement}\label{sec:problemstatement}
\subsection{Notation}
Let $\mathbb{Z}_+$ and $\mathbb{R}_+$ denote the positive integers and reals. For $n\in \mathbb{Z}_+$, we denote $[n] = \{1,2,\ldots,n\}$ and the length-$n$ vector $x^n = (x_1,x_2,\ldots x_n)$. 

\subsection{Channel Model}
A DMC is defined by the single-letter channel transition kernel $P_{Y\vert X} \colon \mc{X} \to \mc{Y}$, where $
\mc{X}$ and $\mc{Y}$ are the input and output alphabets. The DMC acts on each input symbol independently of others, i.e., $P_{Y^n|X^n}(y^n | x^n) = \prod_{i=1}^n P_{Y|X}(y_i | x_i)$ for all $x^n \in \mc{X}^n$ and $y^n \in \mc{Y}^n$.

    \subsection{VLBF Codes}
    We here formalize VLBF codes, which form a subset of VLF codes with a limited number of feedback instances.
    \begin{definition} \label{def:VLF}
        Fix $\epsilon \in (0, 1)$, $N\in\mathbb{R}_+$, and positive integers $M$, $n_1 < \ldots < n_L$. An $(N, L, M, \epsilon)$-VLBF code comprises
        \begin{enumerate}[leftmargin=*]
	\item a common randomness random variable $U$ that has a finite alphabet $\mathcal{U}$ and an associated probability distribution $P_U$, (The realization $u$ of $U$ is revealed to the transmitter and receiver before the start of transmission to initialize the codebook.)
 
	\item a sequence of encoding functions $\mathsf{f}_n\colon \mc{U} \times [M] \times \mc{Y}^{h(n-1)}\to \mc{X}$, $n = 1, \ldots, n_L$, where $h(n) \triangleq \max\{t \in \{0, n_1, \dots, n_L\}\colon t \leq n\}$, generating the channel inputs
 \begin{align}
    X_n = \mathsf{f}_n\left(U, W, {Y}^{h(n-1)}\right), \label{eq:VLFX}
 \end{align}
	\item a random stopping time $\tau \in \{n_1, \dots, n_L\}$ of the filtration generated by $\{U, Y^{n_i}\}_{i = 1}^L$, which satisfies the expected decoding time constraint
	\begin{align}
	    \E{\tau} \leq N,
        \label{eq:averagetau}
	\end{align}
	\item and $L$ decoding functions
	$\mathsf{g}_{n_i} \colon \mc{U} \times \mc{Y}^{n_i} \to [M]$ for $i \in [L]$, giving the estimated message 
 \begin{align}
     \hat{W} = \mathsf{g}_{\tau}(U, Y^{\tau}).
 \end{align} 
When the message $W$ is generated equiprobably on $[M]$, the average error probability must be bounded as
   \begin{align}
        \Prob{\hat{W} \neq W} \leq \epsilon. \label{eq:averageerror}
    \end{align}
	\end{enumerate}
	\end{definition}
	\vspace{-0.1em}
Definition \ref{def:VLF} modifies \cite[Def.~1]{polyanskiy2011feedback} to restrict the number of feedback instances by $L$. VLSF codes with $L$ decoding times defined in \cite{yavas2021VLF} replace \eqref{eq:VLFX} by
     \begin{align}
        X_n = \mathsf{f}_n(U, W);
     \end{align}
VLSF codes are therefore a subset of VLBF codes with feedback used only to decide the stopping time~$\tau$.

	\section{Non-Asymptotic Achievability Bound for VLBF Codes}\label{sec:main}
Before presenting, we recall the random-coding union (RCU) bound \cite[Th.~17]{polyanskiy2010Channel} as we apply it to bound the error probability in the communication phase: for a DMC $P_{Y\vert X} \colon \mc{X} \to \mc{Y}$ and for an arbitrary input distribution $P_{X^n}$ on $\mc{X}^n$, the minimum achievable error probability for fixed-length codes without feedback is upper-bounded by 
\begin{align}
    &\mathrm{rcu}(n, M) \triangleq \label{eq:rcu}\\
    &\mathbb{E}[\text{min}\{1,(M-1)\mathbb{P}[\iota(\bar{X}^n;Y^n)\geq\iota(X^n;Y^n)\vert X^n,Y^n]\}],\notag 
\end{align}
where $n$ is the blocklength, $M$ is the codebook size, $\iota(x; y) \triangleq \log \frac{P_{Y|X}(y|x)}{P_Y(y)}$ is the information density, and 
$(X^n, \bar{X}^n, Y^n)$ is distributed according to $P_{X^n \bar{X}^n Y^n}(x^n, \bar{x}^n, y^n)$ $ = P_{X^n}(x^n) P_{X^n}(\bar{x}^n) P_{Y|X}^n(y^n|x^n)$. The RCU bound follows by analyzing the performance of a random encoder and the maximum likelihood decoder.

Since the receiver in our VLBF code runs a binary hypothesis test to decode the control sequence in the confirmation phase, we formalize binary hypothesis testing next. Let $H_{\mathsf{A}}\colon X \sim P_{\mathsf{A}}$ and $H_{\mathsf{R}}\colon X \sim P_{\mathsf{R}}$ be two hypotheses for the distribution of $X$, where $P_{\mathsf{A}}$ and $P_{\mathsf{R}}$ are distributions on a common alphabet $\mc{X}$. A randomized binary hypothesis test is defined as the probability transition kernel $P_{B|X} \colon \mc{X} \to \{\mathsf{A}, \mathsf{R}\}$, where $\mathsf{A}$ (``Accept") and $\mathsf{R}$ (``Reject") indicate that the test chooses $H_{\mathsf{A}}$ and $H_{\mathsf{R}}$, respectively. The minimum type-II error probability compatible with type-I error probability $\epsilon$ is defined as 
\begin{align}
    \beta_{\epsilon}(P_{\mathsf{A}} \| P_{\mathsf{R}}) \triangleq \min_{\Prob{B = \mathsf{R} | H_{\mathsf{A}}} \leq \epsilon} \Prob{B = \mathsf{A} | H_{\mathsf{R}}},
\end{align}
which is achieved by the log-likelihood ratio test. 

Our main result is an achievability bound for VLBF codes.
 
\begin{theorem}\label{thm:VLBF_main}
Let $L \geq 3$ be an odd number. Fix a DMC $P_{Y|X} \colon \mc{X}\rightarrow\mc{Y}$, a distribution $P_X$ on $\mc{X}$, control symbols $x_{\mathsf{A}}, x_{\mathsf{R}} \in \mathcal{X}$, positive integers $M$ and $n_1 < \dots < n_L$, and type-I error probabilities $\epsilon^{(i)} \in (0, 1)$ for $i \in [\frac{L-1}{2}]$. There exists an $(N, L, M, \epsilon)$-VLBF code with
\begin{align}
    N &\leq n_2 + \sum_{j=1}^{\frac{L-1}{2}} \left[\left(n_{2j +2} - n_{2j} \right) \left(\mathrm{rcu} \left(\sum_{i=1}^j n_{2i-1}- n_{2i-2},M \right) \right.\right. \notag \\
    &\quad \left(1-\beta^{(j)}\right)  + \epsilon^{(j)} \Bigg) \prod_{i=0}^{j-1}p^{(i)} \Bigg] \label{eq:time_bound}\\ 
    \epsilon &\leq\sum_{j=1}^{\frac{L-1}{2}}\mathrm{rcu}\left(\normalsize\sum_{i=1}^j n_{2i-1}- n_{2i-2},M\right)\normalsize \beta^{(j)} \prod_{i=0}^{j-1} p^{(i)} \normalsize \notag \\
    &\quad + \mathrm{rcu}\left(\normalsize\sum_{i=1}^{\frac{L+1}{2}} n_{2i-1}- n_{2i-2},M\right)\normalsize\prod_{i=0}^{\frac{L-1}{2}} p^{(i)}, \label{eq:error_bound}
\end{align}
where $n_0 = 0$, $n_{L+1} = n_L$, and
\begin{align}
\beta^{(i)} &\triangleq \beta_{\epsilon^{(i)}} \left(P_{Y|X = x_{\mathsf{A}}}^{n_{2i}-n_{2i - 1}} \| P_{Y|X = x_{\mathsf{R}}}^{n_{2i}-n_{2i - 1}}\right), \, i \in \left[\frac{L-1}{2} \right] \label{eq:betadef}\\
    p^{(i)} &\triangleq \begin{cases}
    \max\{
\epsilon^{(i)}, 1-\beta^{(i)}\} &\text{if } i \in [\frac{L-1}{2}] \\
    1 &\text{if } i = 0.
    \end{cases}
\end{align}
\end{theorem} 
\begin{IEEEproof}
\thmref{thm:VLBF_main} is derived by constructing a code inspired by the Yamamoto--Itoh scheme \cite{YamamotoItoh} and analyzing its non-asymptotic performance. As in \cite{YamamotoItoh}, the coding scheme that is employed here alternates between communication and confirmation phases. 
The code has $k = \frac{L-1}{2}$ confirmation and $k + 1$ communication phases. To simplify the notation, we define 
\begin{align}
t_i &\triangleq n_{2i - 1}-n_{2i - 2}, \quad i \in [k+1]\\
t_i' &\triangleq n_{2i}-n_{2i - 1}, \quad i \in [k],
\end{align}
where $n_0 = 0, t_{k+1}' = 0$, and $t_i$ and $t_i'$ denote the length of $i$-th communication and $i$-th confirmation phases, respectively.

We generate $M$ independent and identically distributed codewords of length $\sum_{j = 1}^{k+1} t_j$ distributed according to $P_X^{\sum_{j = 1}^{k+1} t_j}$. During $i$-th communication phase, the transmitter emits a subcodeword of length $t_i$ that contains the symbols of the codeword from indices $\sum_{j = 1}^{i-1} t_j + 1$ to $\sum_{j = 1}^{i} t_j$. 
At the end of $i$-th communication phase, the receiver estimates the message using a maximum likelihood rule on $Y^{\sum_{j = 1}^{i} t_j}$. This estimate is sent to the transmitter via a noiseless feedback link. During $i$-th confirmation phase, the transmitter emits either $(x_{\mathsf{A}}, x_{\mathsf{A}}, \dots, x_{\mathsf{A}})$ or $(x_{\mathsf{R}}, x_{\mathsf{R}}, \dots, x_{\mathsf{R}})$, a control sequence of length $t_i'$ telling the receiver whether it should accept or reject its current estimate.
At the end of $i$-th confirmation phase, the receiver constructs the hypothesis test
\begin{align}
H_{\mathsf{A}}^{(i)} \colon Y^{t_i'} \sim P_{Y|X = x_{\mathsf{A}}}^{t_i'}, \quad
H_{\mathsf{R}}^{(i)} \colon Y^{t_i'} \sim P_{Y|X = x_{\mathsf{R}}}^{t_i'},
\end{align}
and applies a log-likelihood ratio test to decide between hypotheses $H_{\mathsf{A}}^{(i)}$ and $H_{\mathsf{R}}^{(i)}$. If $H_{\mathsf{A}}^{(i)}$ is declared, then the receiver accepts its current estimate and informs the transmitter that the communication is over; if $H_{\mathsf{R}}^{(i)}$ is declared, the receiver rejects its current estimate and the transmitter starts the next communication phase. If $(k+1)$-th communication phase is reached, then the receiver accepts its current estimate without further confirmation.

\emph{Error probability analysis:} By construction, the random decoding time in \eqref{eq:averagetau} satisfies $\tau \in \{n_2, n_4, \dots, n_{2k}, n_{2k + 1}\}$, and we take $n_{2k +2} = n_{2k + 1}$ for brevity. We have
\begin{align}
\tau = \begin{cases}
n_{2i} &\text{if } H_{\mathsf{R}}^{(1)}, \dots, H_{\mathsf{R}}^{(i-1)}, H_{\mathsf{A}}^{(i)} \text{ are declared} \\
&\text{ for } i \leq k \\
n_{2k + 2} &\text{if } H_{\mathsf{R}}^{(1)}, \dots, H_{\mathsf{R}}^{(k)} \text{ are declared.}
\end{cases}
\end{align}
Denote the output of the maximum likelihood decoder at the end of $i$-th communication phase by $\hat{W}^{(i)}$ for $i \in [k+1]$. Then, the overall estimate $\hat{W}$ equals $\hat{W}^{(i)}$ if and only if $\tau = n_{2i}$ for $i \in [k + 1]$.
Define the error events
\begin{align}
\mathcal{E}_i &\triangleq \{\hat{W}^{(i)} \neq W \text{ and } \tau = n_{2i} \}, \quad i \in [k+1].
\end{align} 

By the union bound, we have
\begin{align}
\Prob{\hat{W} \neq W} \leq \sum_{j = 1}^{k+1} \Prob{\mathcal{E}_j}. \label{eq:probW}
\end{align}
The terms in the right side of \eqref{eq:probW} are expressed as
\begin{align}
\Prob{\mathcal{E}_j} &= \Prob{\hat{W}^{(j)} \neq W} \Prob{\tau = n_{2j} \middle| \hat{W}^{(j)} \neq W}. \label{eq:bayesterm}
\end{align}
The factors in the right side of \eqref{eq:bayesterm} are bounded as
\begin{align}
    &\Prob{\hat{W}^{(j)} \neq W} \leq \mathrm{rcu}\left(\sum_{i = 1}^j t_i,M\right), \label{eq:rcub} \\
    &\Prob{\tau = n_{2j} \middle | \hat{W}^{(j)} \neq W} \notag \\
    &= \Prob{\text{Declare } H_{\mathsf{R}}^{(1)}, \dots H_{\mathsf{R}}^{(j-1)}, H_{\mathsf{A}}^{(j)} \middle| H_{\mathsf{R}}^{(j)}} \label{eq:tauhypo}  \\
    &= \sum_{b_1 \in \{\mathsf{A}, \mathsf{R}\}} \cdots \sum_{b_{j-1} \in \{\mathsf{A}, \mathsf{R}\}} \Prob{H_{b_1}^{(1)}, \dots, H_{b_{j-1}}^{(j-1)} \text{ are true} \middle | H_{\mathsf{R}}^{(j)}} \notag \\
    &\quad \Prob{\text{Declare } H_{\mathsf{R}}^{(1)}, \dots H_{\mathsf{R}}^{(j-1)}, H_{\mathsf{A}}^{(j)} \middle| H_{b_1}^{(1)}, \dots, H_{b_{j-1}}^{(j-1)}, H_{\mathsf{R}}^{(j)}} \label{eq:condition} \\
    &\leq \max_{b_\ell \in \{\mathsf{A}, \mathsf{R}\}, \ell \in [j-1]} \mathbb{P} \big[\text{Declare } H_{\mathsf{R}}^{(1)}, \dots H_{\mathsf{R}}^{(j-1)}, H_{\mathsf{A}}^{(j)} \notag \\
    &\quad \quad \mid H_{b_1}^{(1)}, \dots, H_{b_{j-1}}^{(j-1)}, H_{\mathsf{R}}^{(j)} \big] \\
    &= \beta^{(j)} \prod_{i=0}^{j-1} p^{(i)}, \label{eq:betaj}
\end{align}
where \eqref{eq:rcub} follows from the RCU bound \eqref{eq:rcu}, and \eqref{eq:tauhypo} follows since given the event $\{\hat{W}^{(j)} \neq W\}$, $\tau = n_{2j}$ occurs if and only if the first $j-1$ tests reject the null hypothesis, and \mbox{$j$-th} test erroneously accepts it. Whether $H_{\mathsf{A}}^{(i)}$ or $H_{\mathsf{R}}^{(i)}$ is true is a random variable depending on the transmission until \mbox{$i$-th} communication phase. Equality \eqref{eq:condition} conditions on $2^{j-1}$ combinations of true hypotheses in the first $j-1$ confirmation phases, and \eqref{eq:betaj} uses the fact that declaring $\mathsf{A}$ or $\mathsf{R}$ in the $i$-th test is conditionally independent from the first $i-1$ hypotheses given $H_{b}^{(i)}$, regardless of whether $b\in\{\mathsf{A},\mathsf{R}\}$.  
Combining \eqref{eq:probW}--\eqref{eq:rcub} and \eqref{eq:betaj}, we get \eqref{eq:error_bound}.

\emph{Average decoding time analysis:} Since $\tau \in \{n_{2j} \colon j \in [k +1]\}$, we have
\begin{align}
\E{\tau} = n_2 + \sum_{j = 1}^k (n_{2j + 2} - n_{2j}) \Prob{\tau > n_{2j}}. \label{eq:etau}
\end{align}
By conditioning on whether the event $\{\hat{W}^{(j)} \neq W\}$ occurs, we bound the probability terms in \eqref{eq:etau} as
\begin{align}
&\Prob{\tau > n_{2j}} \notag \\
&\leq \Prob{\hat{W}^{(j)} = W} \Prob{\text{Declare } H_{\mathsf{R}}^{(1)}, \dots, H_{\mathsf{R}}^{(j)} \middle| H_{\mathsf{A}}^{(j)}}  \notag \\ 
&+ \Prob{\hat{W}^{(j)} \neq W} \Prob{\text{Declare } H_{\mathsf{R}}^{(1)}, \dots, H_{\mathsf{R}}^{(j)} \middle| H_{\mathsf{R}}^{(j)}}. \label{eq:prob4terms} 
\end{align}
We bound $\Prob{\hat{W}^{(j)} = W}$ by 1 and the remaining three probability terms in \eqref{eq:prob4terms} using arguments similar to \eqref{eq:rcub}--\eqref{eq:betaj}, which yields  \eqref{eq:time_bound}.
\end{IEEEproof}

\section{Numerical Evaluation over the BSC} \label{sec:numer}
We evaluate the bounds in \thmref{thm:VLBF_main} over BSC($p$), $p \in (0, 1)$. Under an equiprobable $P_{X^n}$ \cite[Th.~33]{polyanskiy2010Channel}, the RCU error probability bound reduces to
\begin{align}
&\mathrm{rcu}(n, M) \\
&= \sum_{k=0}^n {n\choose k} p^k (1-p)^{n-k} \min\left\{1,\left(M-1\right)\sum_{j=0}^k {n \choose j}2^{-n}\right\}. \label{eq:rcuBSC} \notag 
\end{align}

In \cite{YamamotoItoh}, in order to optimize the error exponent of type-II error probability, the ``$\mathsf{A}$" and ``$\mathsf{R}$" symbols transmitted in the confirmation phase are chosen as
\begin{align}
(x_{\mathsf{A}}, x_{\mathsf{R}}) = \argmax_{x_{\mathsf{A}}, x_{\mathsf{R}} \in \mc{X}} D\left(P_{Y|X = x_{\mathsf{A}}} \| P_{Y|X = x_{\mathsf{R}}}\right).
\end{align}
Accordingly, we set $(x_{\mathsf{A}}, x_{\mathsf{R}}) = (1, 0)$. By the Neyman-Pearson lemma (e.g., \cite[Lemma~57]{polyanskiy2010Channel}), for the BSC$(p)$, $\beta^{(i)}$ in \eqref{eq:betadef} is given by
\begin{align}
\beta^{(i)} = \Prob{T_i > \gamma_i} + (1-\lambda_i) \Prob{T_i = \gamma_i},
\end{align}
where the fixed type-I error probability is parameterized as
\begin{align}
\epsilon^{(i)} = \Prob{Z_i < \gamma_i} + \lambda_i \Prob{Z_i = \gamma_i}, \label{eq:alpha}
\end{align}
$T_i \sim \mathrm{Binom}(t_i', p)$, $Z_i \sim \mathrm{Binom}(t_i', 1-p)$, and $(\gamma_i, \lambda_i)$ is the unique solution to \eqref{eq:alpha} with $\lambda_i \in [0, 1)$.

Using \eqref{eq:rcuBSC}--\eqref{eq:alpha}, we compare our VLBF achievability bound in Theorem $\ref{thm:VLBF_main}$ to the VLSF achievability bound presented in \cite[Th.~3]{yavas2021VLF} for BSC(0.11). For a given codebook size $M$, average error probability tolerance $\epsilon$, and number of feedback instances $L$, we perform numerical optimization over the possible feedback times and hypothesis testing parameters $\gamma_i$ in \eqref{eq:alpha} ($\lambda_i$'s are set to 0 since their effect is negligible) to minimize the corresponding expected decoding time for both bounds. Parameters are optimized via brute force for $L=3$. For $L=5$, due to the increased number of parameters, we use the stochastic response surface method \cite{Regis}, which applies to optimize computationally expensive objective functions where derivatives are unavailable.

\begin{figure}[!htbp]
            \centering
            \includegraphics[width=0.48\textwidth]{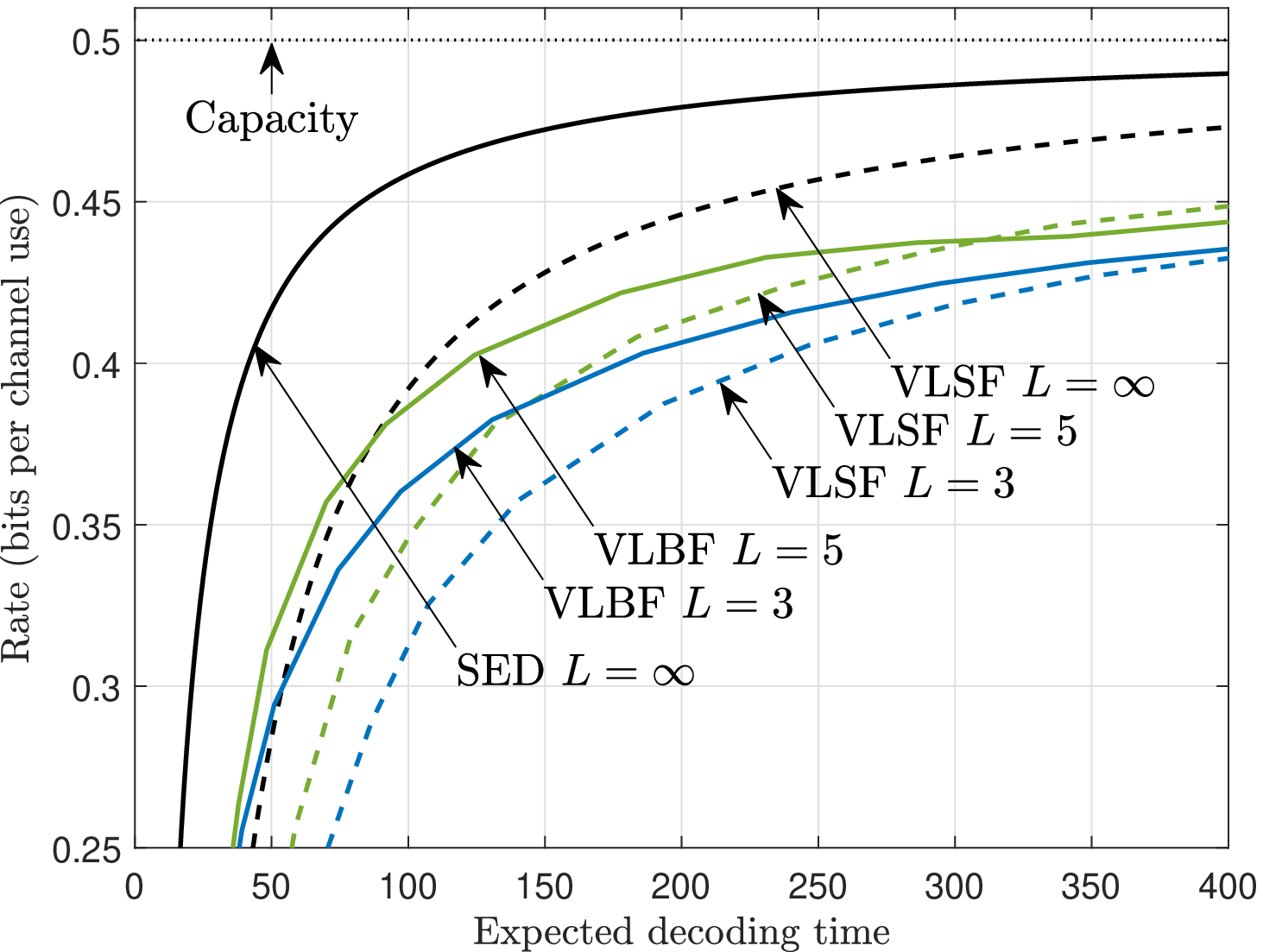}
            
            \vspace{0.3cm}
            
            \includegraphics[width=0.48\textwidth]{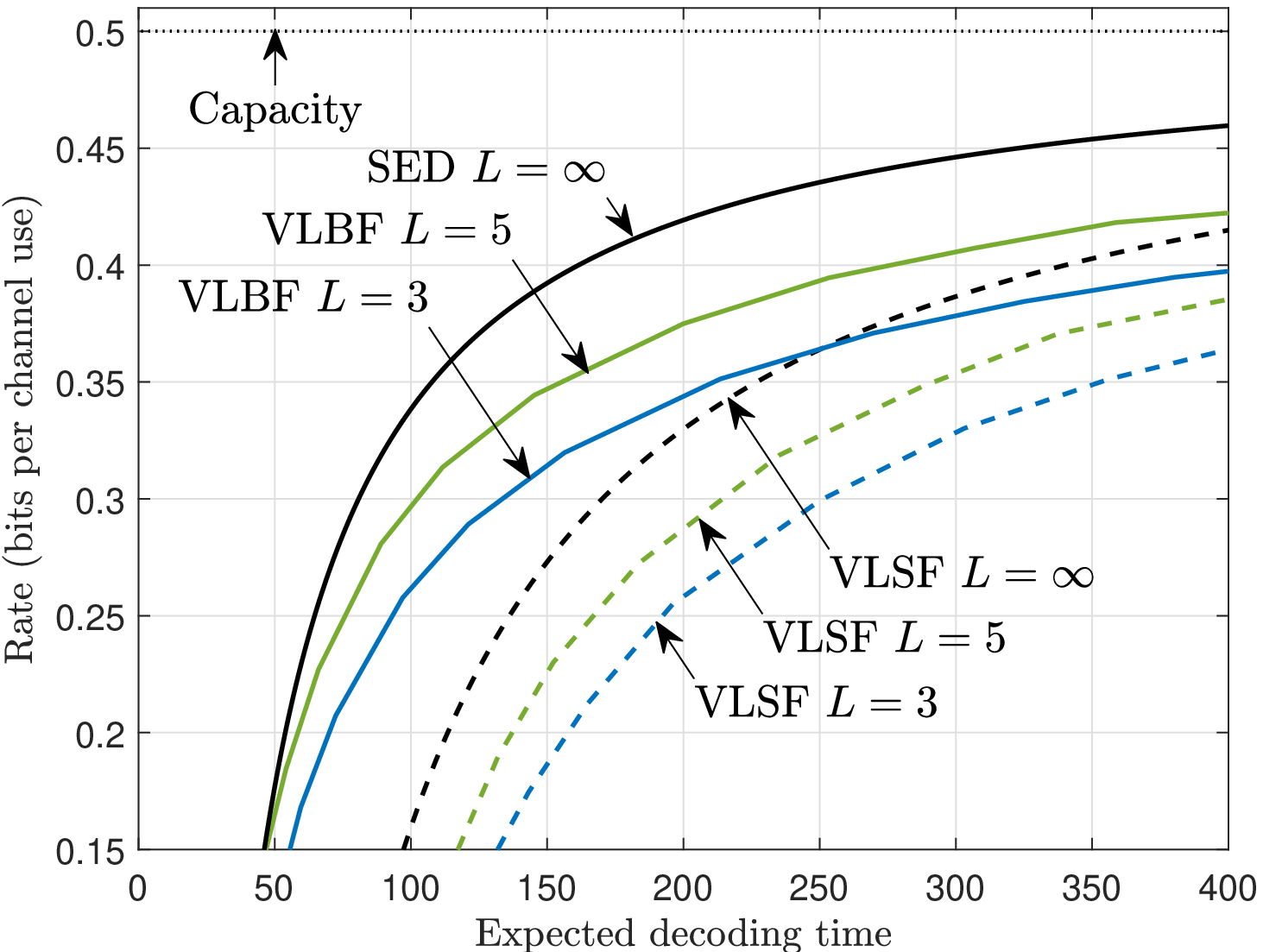} 
            \caption{Numerical evaluations of  \thmref{thm:VLBF_main} over BSC(0.11) for $L \in \{3,5\}$. The corresponding error probability constraints are $10^{-3}$ (top) and $ 10^{-10}$ (bottom). Rate is calculated as $\frac{\log_2 M}{N}$ for  $(N, L, M, \epsilon)$-VLBF codes.}
            \label{fig:rate_time_plots}
        \end{figure}

\figref{fig:rate_time_plots} shows numerical evaluations of the achievability bounds over BSC(0.11). Our VLBF code outperforms VLSF codes with the same number of feedback instances at shorter blocklengths, especially when the error probability is lowered to $10^{-10}$. In the regime where expected decoding time $N \leq 400$ and error probability $\epsilon = 10^{-10}$, our VLBF code with $L=5$ feedback instances achieves higher rates than the non-asymptotic achievability bound presented in \cite[Th.~2]{polyanskiy2011feedback} for VLSF codes with $L=\infty$. Our achievability bound for VLBF $L=5$ is outperformed by its stop-feedback counterpart for $N > 310$, indicating room for improvement in how our code uses available feedback opportunities. We also compare our VLBF code with the achievability bound given by \cite[Th.~7]{yang2022SED} for the SED encoder, which uses dense feedback with $L=\infty$ and thus outperforms all the other VLF codes shown.

\begin{figure}
            \centering
            
            \includegraphics[width=0.47\textwidth]{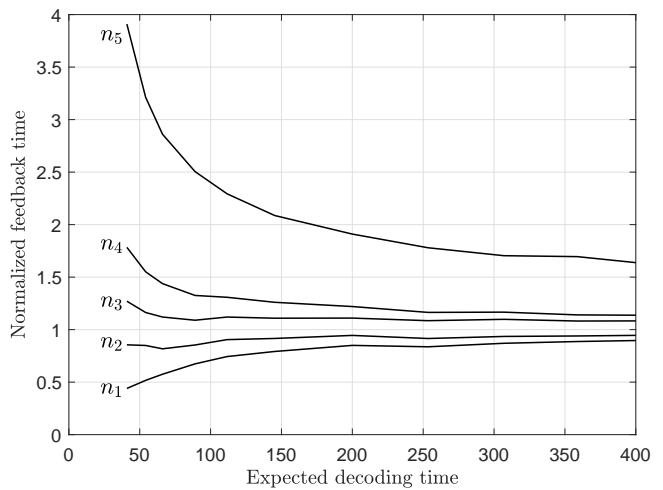}
            \caption{Optimized feedback times $\{n_i\}_{i = 1}^L$ in \thmref{thm:VLBF_main} normalized by the expected decoding time $N$ for $\epsilon = 10^{-10}$ and $L = 5$. }
            \label{fig:normalized_feedback_plots}         
        \end{figure}

We show the optimized feedback times $n_i$ in \figref{fig:normalized_feedback_plots}. For $L=5$, the optimal decoding times are not periodic as in~\cite{YamamotoItoh}. Instead, as the expected decoding time increases, the confirmation phases become shorter relative to the communication phases as seen from the diminishing gaps $n_2-n_1$ and $n_4-n_3$. The normalized feedback times $n_1$ to $n_4$ get closer to the expected decoding time, while the gap between $n_4$ and $n_5$ remains relatively large compared to the other gaps. This is because the term that contributes to the error probability is governed by the type-II error probability of the confirmation phases; the lack of a confirmation phase after the last communication phase is compensated by a longer final communication phase.


        \section{Conclusion} \label{sec:conclusion}
        In this paper, we formalize the notion of VLBF codes that fix the number of feedback bursts while leaving the actual contents of each feedback burst unconstrained. We derive the first VLBF achievability bound and optimize the times at which feedback bursts occur over the BSC. At short blocklengths, our VLBF codes strongly outperform VLSF codes with the same number of feedback instances and at low error tolerances even beat the achievability bound given by Polyanskiy \emph{et al.} \cite{polyanskiy2011feedback} for VLSF codes with $L=\infty$. Our results demonstrate the benefits of utilizing richer feedback over sparse feedback and that limiting feedback to sparse bursts does not significantly hamper performance.

         One direction for further work is to extend \thmref{thm:VLBF_main} to the Gaussian channel with maximal power constraints, where we would need to modify Shannon's bound in \cite{shannon1959Probability} (see also \cite[eq. (42)]{polyanskiy2010Channel}) to guarantee that the power constraints are satisfied for all $L$ decoding times. The code structure in \cite{yavas2021VLF} where each codeword is a concatenation of $L$ subcodewords drawn independently from a uniform distribution on their respective high-dimensional spheres can be used for this purpose. Another potential future direction is to derive a tight non-asymptotic converse bound for VLBF codes. While the converse bound in \cite[Th.~4]{polyanskiy2011feedback} for $L = \infty$ applies, it gives a rate strictly above the capacity. Leveraging the fact that $L$ is finite in the converse proof is a challenging task since the data-processing inequality used in the proof of \cite[Th.~4]{polyanskiy2011feedback} disallows it.
        
\clearpage
\bibliographystyle{IEEEtran}
\IEEEtriggeratref{15}
\bibliography{mac}

\end{document}